\documentclass[fleqn,usenatbib]{mnras}

\usepackage{newtxtext,newtxmath}
\usepackage[T1]{fontenc}

\DeclareRobustCommand{\VAN}[3]{#2}
\let\VANthebibliography\thebibliography
\def\thebibliography{\DeclareRobustCommand{\VAN}[3]{##3}\VANthebibliography}

\usepackage{graphicx}	% Including figure files
\usepackage{amsmath}	% Advanced maths commands
\usepackage{booktabs}
\usepackage{tabularx}
\usepackage[table,x11names]{xcolor}
\usepackage{subcaption}
\usepackage{appendix}

\title[Massive Compact Binaries]{Mapping Progenitors of Binary Black Holes and Neutron Stars with Binary Population Synthesis}

\author[Weller et al.]{
Miqaela K. Weller,$^{1}$\thanks{E-mail: weller.133@osu.edu}
Jennifer A. Johnson,$^{1, 2}$
\\
$^{1}$Department of Astronomy, The Ohio State University, 140 West 18th Avenue, Columbus, OH 43210, USA\\
$^{2}$Center for Cosmology and AstroParticle Physics, The Ohio State University Columbus, OH 43210, USA
}

\date{Accepted XXX. Received YYY; in original form ZZZ}

\pubyear{2022}

\begin{document}
\label{firstpage}
\pagerange{\pageref{firstpage}--\pageref{lastpage}}
\maketitle

\begin{abstract}
The first directly observed gravitational wave event, GW150914, featuring the merger of two massive black holes, highlighted the need to determine how these systems of compact remnant binaries are formed. We use the binary population synthesis code COSMIC (Compact Object Synthesis and Monte Carlo Investigation Code) to predict the types of massive stars that will show significant radial velocity variations, indicative of a potential compact object (i.e. a black hole or neutron star) orbiting the star. We “observe” the binaries generated in the populations with a similar number of epochs and RV accuracy as planned for the Milky Way Mapper. In this analysis, we are especially interested in systems where a compact remnant is orbiting a massive O or B star as these systems survived the first supernova and neutron star kick. We test the ability of the Milky Way Mapper observing strategy to distinguish among different mass loss and kick prescriptions. We find that Wolf-Rayet stars or hot subdwarfs in binaries could be detectable (i.e. luminous, high $\Delta RV_{max}$), viable progenitors of such objects, while the different prescriptions primarily affect the number of sources. 
\end{abstract}

\begin{keywords}
gravitational waves -- (transients:) black hole mergers -- (transients:) black hole - neutron star mergers -- (transients:) neutron star mergers -- binaries: general 
\end{keywords}

%%%%%%%%%%%%%%%%%%%%%%%%%%%%%%%%%%%%%%%%%%%%%%%%%%

%%%%%%%%%%%%%%%%% BODY OF PAPER %%%%%%%%%%%%%%%%%%

\newcommand{\comment}[1]{}

\section{Introduction}

In 2015, the Laser Interferometer Gravitational Wave Observatory and Virgo (LIGO/Virgo) directly observed gravitational waves. The gravitational waves were created by a black hole binary merging to form one supermassive black hole (BH) about 62 times the mass of the Sun \citep{Abbott2016}. Such massive binaries are the strongest gravitational wave sources to which LIGO is sensitive to and are inherently distinct signals from electromagnetic radiation. Since this first measurement, more detections have been made, including more BBH (binary black hole) mergers, two BNS (binary neutron star) mergers, and two possible BH+NS mergers \citep[e.g.,][]{Abbott_2019, Abbott_2021, AbbottBHNS}. The two neutron star mergers were further confirmed and localized with electromagnetic observations, including short $\gamma$-ray bursts. \citep{Abbott2017}. So far, one of the surprising discoveries of the gravitational wave events detected is the presence of BH+BH mergers where the combined mass greatly exceeds the remnant masses expected from massive star evolution. In particular, \citet{Abbott_2021} found at least three black holes with masses exceeding 45 M$_{\odot}$. 

The formation of BBH, BNS, and BH+NS systems is therefore of considerable interest for the population of LIGO gravitational wave sources. The formation of these systems in general, including systems that will not merge on cosmically relevant timescales, is an excellent laboratory for studies of binary populations and binary interactions. Most O stars are in binaries and binary interactions such as mass stripping are common \citep{Sana2012}, showing that the co-evolution of two massive stars adds many new mechanisms that could affect the system. Stars can undergo considerable amounts of mass transfer, both stable and unstable. Stable mass transfer can happen through Roche lobe overflow or winds \citep{bse2} while unstable mass transfer can initiate a common envelope phase and expulsion of substantial amounts of material from the system, thus reshaping the dynamics of the system. 
Other mechanisms of co-evolution such as tidal interaction can circularize the orbits of the stars \citep{bse2}. Furthermore, with massive binaries, the threat of disruption in either of the two supernova kicks is also large, as is seen in observations, as well as 50$\%$ mass loss due to the virial theorem. For example, pulsars can have birth velocities, ascribed to asymmetric supernovae collapses, that far exceed the escape velocity of a binary system \citep{Lyne1994}. Additionally, kicks could also bring about the higher fraction of runaway stars seen in higher mass bodies, an indication that a lot of systems do not survive the first supernova. \citep{Gies1986}. Essentially, to even have any binary system with a neutron star or black hole, many features of binary evolution have to be withstood. 

\citet{Abbott_asymmetry} used the existing GW events to constrain the merging compact binary population, including mass distributions and mass ratios. We can approach the problem the other way, by exploring the properties of the progenitor systems. However, converting a mass distribution of living stars to their compact remnants is not straightforward because of the many factors that influence the evolution of massive stars, including mass loss and supernova kicks, e.g. \cite{Giacobbo2018}. While the dynamical formation of double compact binaries by capture is viable in dense stellar environments \citep[e.g.,][]{East_2012}, such systems do not produce visible progenitors. In this paper, we focus on isolated binaries, which are believed to contribute substantial numbers of progenitor systems \citep[e.g.,][]{Bouffanais_2021}.

Historically, finding black holes and neutron stars in a binary system has been accomplished through detection of X-ray emission due to accretion. However, such a signature requires ongoing mass transfer from the stellar companion through either a strong wind or Roche-lobe overflow and is therefore present for a limited time and for a limited subset of binaries \citep{Thompson2019}. Pulsars have also been utilized to detect companions in non-interacting binaries \citep[e.g.,][]{Yan2021, Lorimer_2021, Eatough2021}, through the use of radio timing and the eclipsing of such timings due to a companion. Nonetheless, incorporating other methods of observation, such as the measurement of radial velocity and ellipsoidal variability, we can detect BHs and NSs in binaries that are not interacting and, in some cases, will never interact, with the challenge of separating them from the far larger number of stellar binaries. 

The advent of multi-object spectroscopic stellar surveys has made monitoring stars for radial velocities on an industrial scale a viable prospect. The APOGEE survey \citep{Majewski_2017} observed over 650,000 stars at least three times to identify RV variables, both to eliminate their influence on Galactic kinematic properties \citep{Majewski_2017} and to study them as scientifically interesting objects \citep[e.g.,][]{Troup_2016, Badenes2018}. The LAMOST survey is measuring multi-epoch RVs for $>10^5$ stars at more than one epoch, using both low-resolution \citep[e.g.,][]{Qian_2019} and medium-resolution. Gaia Data Release 2 included $\sim10^7$ average RV measurements \citep{Katz_2019}; with Data Release 3\footnote{https://www.cosmos.esa.int/web/gaia/release}, Gaia will begin releasing data on both astrometric and RV binaries. The Milky Way Mapper survey is planning on multi-epoch observations \citep{Kollmeier_2017}. Sparse RV data in combination with photometric or astrometric information has already been used to identify BH candidates around low mass stars  \citep[e.g.,][]{Gu_2019, Thompson2019, Jayasinghe2021, Jayasinghe2022}, but similar data have not yet been obtained for massive stars in these large scale surveys. Therefore, we have a largely incomplete picture of the stages of binary evolution, such as post common-envelope phases and survival after the first supernova. However, studies have been done to show binary properties (including the binary fraction) at the time of SNe, giving some insight into binary evolution \citep[e.g.,][]{Kochanek2018, Kochanek2019, Kochanek2021}.

In this paper, we study the observability of progenitor systems to BH+BH, BH+NS, and NS+NS systems in few-epoch radial velocity surveys of many stars, with a focus on those that merge within 13.7 billion years. We generate populations of compact binaries using the Compact Object Synthesis and Monte Carlo Investigation Code (COSMIC\footnote{https://github.com/COSMIC-PopSynth/COSMIC} \cite{Breivik2020}). We measure the maximum radial velocity change $\Delta RV_{max}$ throughout the evolutionary history for these systems. \citet{Badenes2018} showed that this statistic provides crucial information on the binary population even if complete orbital information is not recovered. In Section \ref{sec:cosmic}, we describe the steps used to create a binary population from COSMIC and the methods used to calculate the radial velocities to mimic a few-epoch survey. In Section \ref{sec:respop}, we analyze the results of a population built using the default COSMIC parameters (e.g., solar metallicity, iron core-collapse SN kicks drawn from a Maxwellian distribution). In Section \ref{sec:params}, we analyze the difference in the distribution of the $\Delta RV_{max}$ statistic if we change either the metallicity or the kick prescription to illustrate how observations of progenitor systems can help constrain these parameters.

\section{Predicting $\Delta RV_{max}$ for Progenitors of Compact Object Mergers}
\label{sec:cosmic}

COSMIC is a python-based rapid binary synthesis code that can simulate the BH/NS binaries and their progenitors. \citep{Breivik2020}. This code is adapted from \cite{bse1,bse2} and can trace the evolution of binary systems, including the evolution of the individual stars.  In this paper, we used v3.3 of COSMIC to run our populations, and it has since been updated to v3.4. 

COSMIC reports the evolutionary states of the two stars in a binary system based on the \citet{bse1} classifications. We summarize key points here and refer the reader to \citet{bse1} for additional information. On the Main-Sequence (MS), the underlying BSE code distinguishes between mostly or fully convective stars ({\tt MS, $< 0.7 {\mathrm{M}_\odot}$}) and main-sequence stars with appreciable radiative zones ({\tt MS, $> 0.7 {\mathrm{M}_\odot}$}) because this modifies the angular momentum loss. The 'Naked Helium Star' sequence is the predicted position on the H-R diagram of pure helium stars that have formed from massive stars that have large amounts of mass loss, either through winds or binary mass transfer. Observationally, these are Wolf-Rayet stars and their descendants or hot subdwarfs \citep[e.g.,][]{Gotberg2018}.

The default parameters used by COSMIC represent a reasonable set of choices based on our current knowledge, such as using the binary mass ratios, binary fractions, and the initial mass function from \citet{Moe2017}. Therefore, for our initial run, we elect to use the default parameters defined by COSMIC. 

Part of our discussion focuses on the observable radial velocities for systems that will merge in a Hubble time as BH+BH, BH+NS, or NS+NS binaries and therefore possibly detectable by LIGO. While there is a slight inconsistency in our approach of using systems that have merged in a Hubble time to predict the magnitude of radial velocity variations observable now, our goal of understanding the manifestations of such systems should be little affected by this approximation.

%JAJ: put this later!!! The systems that will merge have high RVs during their progenitor stages, so this exploration establishes the minimum RV precision needed to statistically investigate systems that will end up as two massive remnants.

\subsection{Simulations}
\label{sec:sims} 

We start by generating a binary population using the multidimensional sampler in which COSMIC incorporates parameter distributions given by \cite{Moe2017}. We set the size of the population to be 100,000 and specify the final star types to be neutron stars or black holes. For the initial binary population, we also use a user-specified star formation history (default):
\begin{itemize}
    \item SF\_start =13700.0 
    \item SF\_duration = 0.0 
    \item Z = 0.02 
\end{itemize}
where the start of star formation begins at the time of the Big Bang, lasts for 0.0 Myr, with a binary population metallicity of 0.02. All systems are then followed for a Hubble Time. Since we focus on systems with two massive short-lived stars, treating our population as a burst is sufficient to address the questions of the expected RV variability as a function of evolutionary state for massive star binaries.

Once our initial binary population is generated, we evolve the binaries using COSMIC’s evolve class, using the default parameters given by 
COSMIC. The first line of Table \ref{tab:kicksims} summarizes the outcomes of our population for all 100,000 binary systems.

\subsection{Calculating $\Delta RV_{max}$}
\label{sec:methods}

From the output of the simulation, we receive certain properties of each star and binary system, such as the stellar evolutionary phase and orbital parameters, at timesteps when key evolutionary changes occur, such as the beginning or end of Roche Lobe Overflow (RLOF). These timesteps are therefore not equally spaced. Table \ref{tab:bpp} gives an example of the output of COSMIC for one binary system of our population, which illustrates how COSMIC defines important timesteps. Between each timestep the orbital properties of the system still change, but we only construct a single radial velocity curve for each output timestep. We use the masses of the stars, the eccentricity of the system, the orbital period, and the semi-major axis to calculate the radial velocity curve in two cases:
\begin{enumerate}
    \item RV of the primary star when there are two massive living stars. We assume that when there are two living stars that the more massive star will outshine the lower mass star and RV variability of the primary will be detectable in a spectrum of the system. 
    \item RV of the secondary star when there is one living star and one compact remnant (the primary), where there is only one source with detectable RV variability. 

\end{enumerate}
We start by calculating the semi-amplitude, K, in each of our two cases:

\begin{eqnarray}
    K_{1} = \frac{2 \pi a m_{2} \sin{i}}{P (m_{1}+m_{2}) \sqrt{1 - e^{2}}}
    \label{eq:K1}
\end{eqnarray}

\begin{eqnarray}
    K_{2} = \frac{2 \pi a m_{1} \sin{i}}{P (m_{1}+m_{2}) \sqrt{1 - e^{2}}}
    \label{eq:K2}
\end{eqnarray}
where the inclination angle $i$ is initially set through a pseudo-random number generator (drawn from a cos$i$ distribution). Note that throughout the evolution of the system, the primary remains labeled as the primary even after it becomes a compact remnant and is no longer the more massive body. 

The calculation of the radial velocity as a function of time is more involved to avoid requiring that the eccentricity is small. The basic equation is

\begin{eqnarray}
    V_{\rm rad} = K (\cos{(\omega + v)} + e \cos{\omega})
    \label{eq:RV}.
\end{eqnarray}
$\omega$ is another angle that is randomly generated, but we must also calculate the angle $v$, the true anomaly, in such a way that we can input specific epochs and get the radial velocity of the system. We accomplish this by first calculating the mean anomaly, which is where the time dependence occurs.

\begin{eqnarray}
    M = \frac{2 \pi t}{P}
    \label{eq:mean}
\end{eqnarray}
From here, we can calculate the eccentric anomaly using Kepler's Equation.

\begin{eqnarray}
    M = E - e \sin{E}
    \label{eq:kepler}
\end{eqnarray}
This equation, however, is a transcendental equation and cannot be solved without numerical analysis. As a result, we employ the Newton-Raphson method to numerically calculate the root of this equation after calculating the mean anomaly $M$, equation \ref{eq:mean}, with a given epoch.

\begin{figure}
	\includegraphics[width=\columnwidth]{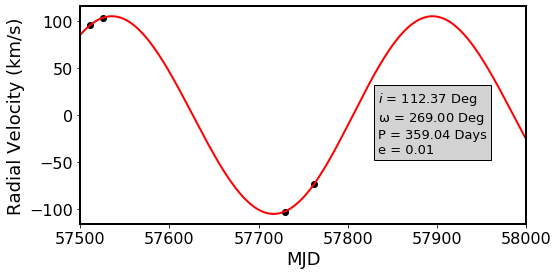}
    \caption{The solid red line represents the radial velocity curve for the highlighted timestep in Table \ref{tab:bpp} after $4.2271$ Myr of evolution. The primary star is now a BH ($kstar\_1 =  14$) while the secondary is still a main sequence star ($kstar\_2 = 1$). The inclination angle is $i = 112.37$\textdegree  and eccentricity is near zero, but not exact. The black points represent our random "observations," used to calculate $\Delta RV_{max}$.}
    \label{fig:rvsamp}
\end{figure}

We next simulate the number of observations and their cadence for a few-epoch RV survey. For each distinct orbit of a given system, we randomly select a starting epoch as well as additional epochs both within a few days and months to years apart utilizing random number generators. This scheme was based on the typical cadences from the APOGEE survey. We select an arbitrary time and add observations based on random draws between 1 - 20 days and 20 - 500 days. We elected to gather 3-4 observations only, to match the number of observations in few-epoch surveys \citep[e.g.,][]{Badenes2018}. Figure \ref{fig:rvsamp} shows an example of the sampled RV curve of the system in Table \ref{tab:bpp} at the highlighted timestep and the spacing of our 3-4 observations. 

\begin{table*}{BPP Example Output from COSMIC}
    \centering
    % Need to figure out how to make it textwidth
    \caption{Example output from COSMIC for parameters we use in this paper for a single binary system in the population that eneded with the merging of two black holes. Many more parameters are present in the output of the BPP dataframe in COSMIC. }
	\label{tab:bpp}
    \begin{tabularx}{0.8\textwidth}{c c c c c c c c c c}
    \toprule
tphys & mass\_1 & mass\_2 & kstar\_1 & kstar\_2 & sep & porb & ecc & evol\_type & bin\_num \\ \midrule
0 & 84.0026 & 54.0964 & 1 & 1 & 540.2872 & 123.8551 & 0.5565 & 1 & 26268 \\
3.6172 & 48.1447 & 44.2675 & 2 & 1 & 777.1501 & 261.1950 & 0.5493 & 2 & 26268 \\
3.6184 & 48.0838 & 44.2693 & 2 & 1 & 350.4608 & 79.1235 & 0 & 3 & 26268 \\
3.6219 & 33.9440 & 58.0405 & 4 & 1 & 381.2995 & 89.9734 & 0 & 2 & 26268 \\
3.6291 & 18.5094 & 72.9804 & 4 & 1 & 728.6754 & 238.3346 & 0 & 4 & 26268 \\
3.6301 & 18.4170 & 72.9962 & 7 & 1 & 728.7640 & 238.4778 & 0 & 2 & 26268 \\
4.1944 & 8.4305 & 67.2429 & 8 & 1 & 880.0229 & 347.8096 & 0 & 2 & 26268 \\
4.2271 & 8.0506 & 66.9322 & 8 & 1 & 888.1171 & 354.2392 & 0 & 15 & 26268 \\ \rowcolor{lightgray}
4.2271 & 7.5506 & 66.9322 & 14 & 1 & 894.1193 & 359.0355 & 0.0067 & 2 & 26268 \\
5.2997 & 7.5580 & 49.9486 & 14 & 2 & 1155.2845 & 600.1317 & 0.0067 & 2 & 26268 \\
5.3026 & 7.5598 & 49.6157 & 14 & 2 & 1153.5129 & 600.4832 & 0 & 3 & 26268 \\
5.3026 & 7.5598 & 49.6157 & 14 & 2 & 1153.5129 & 600.4832 & 0 & 7 & 26268 \\
5.3026 & 7.5598 & 19.1422 & 14 & 7 & 5.0758 & 0.2565 & 0 & 8 & 26268 \\
5.3026 & 7.5598 & 19.1422 & 14 & 7 & 5.0758 & 0.2565 & 0 & 4 & 26268 \\
5.8699 & 8.0976 & 8.5117 & 14 & 8 & 6.6323 & 0.4857 & 0 & 2 & 26268 \\
5.9021 & 8.1255 & 8.1286 & 14 & 8 & 6.7308 & 0.5020 & 0 & 16 & 26268 \\
5.9021 & 8.1255 & 7.6286 & 14 & 14 & 6.9514 & 0.5352 & 0.0317 & 2 & 26268 \\
388.3391 & 8.1255 & 7.6286 & 14 & 14 & 8.75E-05 & 2.39E-08 & 0 & 3 & 26268 \\
388.3391 & 8.4305 & 15.7541 & 15 & 14 & 0 & 0 & -1 & 6 & 26268 \\
13700 & 0 & 15.7541 & 15 & 14 & 0 & 0 & -1 & 10 & 26268 \\ \bottomrule
\end{tabularx}
\end{table*}

We characterize the radial velocity variation of the system using the statistic from \citet{Badenes2018}.

\begin{eqnarray}
    \Delta RV_{max} = \rm{max}(RV_{measured}) - \rm{min}(RV_{measured})
    \label{eq:delrv} 
\end{eqnarray}
This equation can be used even for noisy or few-epoch data while still probing the radial velocity curve of a population in a statistical sense. In the example radial velocity curve seen in Figure \ref{fig:rvsamp}, the $\Delta RV_{max}$ measured would be approximately 200 km s$^{-1}$ based on this equation where we have nearly captured the full radial velocity amplitude. For each timestep in COSMIC we calculate a radial velocity curve and a value for $\Delta RV_{max}$. 

\section{Results for the Default Population}
\label{sec:respop}

For the rest of the paper, we first group binary systems in the population based on the final end states because gravitational wave detectors can reliably tell us which type of remnants merged, and then by the evolutionary status of the members at the time of the RV observations. For example, those that end up as BH+BH are further categorized into the group BH+Living Star or into the group Living Star+Living Star. We elect to do such a categorization to connect those end states with early stages that can be detected by RV variability of living stars. The same is applied to BH+NS as well as NS+NS systems.

\begin{figure}
	\includegraphics[width=\columnwidth]{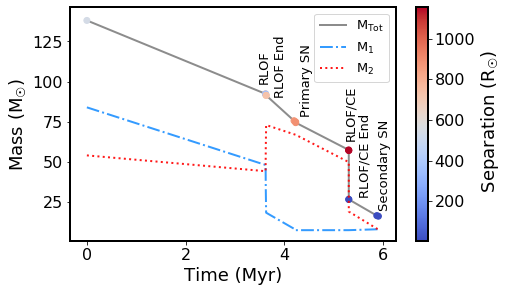}
    \caption{The evolutionary history of a representative merging BH+BH binary system. The grey line shows the total mass of the system up to the death of the secondary while the blue and red lines are the masses of the primary and secondary, respectively. The orbital separation is denoted by the colour bar. Note that 'Primary SN' and 'Secondary SN' refer to the formation of the compact objects; COSMIC only includes this in their documentation even though FeCCSN typically produce NSs (and not BHs). \citep{Sukhbold2016}}
    \label{fig:singleBHBH}
\end{figure}

To illustrate the changes in the binary system that affect its radial velocity, we show in Figure \ref{fig:singleBHBH} the evolutionary history of one system that merges as a BH+BH. Both the primary and secondary experience mass loss, first gradually over the main sequence and then more drastically when RLOF begins. The SN of the primary ejects additional mass from the system and causes the separation to increase markedly. Finally, the onset of the second RLOF phase leads to a common envelope phase and a further decrease in mass accompanied by a large decrease in the separation because of the drag force from the envelope. The secondary becomes a Naked Helium Star as a result of this interaction. The lower total mass and larger separation means that the phase where one member was a BH and the other is not yet evolved to large enough radii to do RLOF has the lowest semi-amplitude. We note that the RLOF and Common Envelope (CE) phases are extremely short-lived and would not be observed as X-ray binaries as a result. 

\subsection{BH+BH Mergers}

\begin{figure*}
\begin{tabularx}{1\textwidth}{cc}
\centering
  \includegraphics[width=85mm]{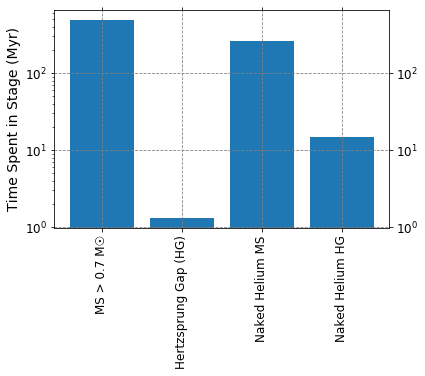} &   \includegraphics[width=85mm]{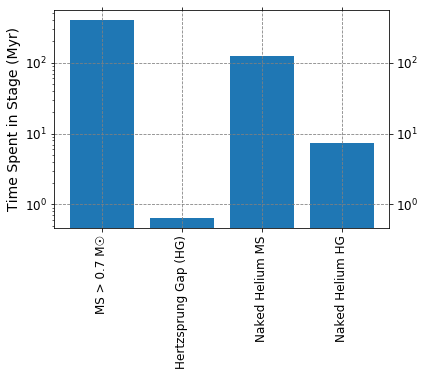} \\
(a) Final State: BH+BH Merger & (b) Final State: BH+NS Merger \\[6pt]
\multicolumn{2}{c}{\includegraphics[width=85mm]{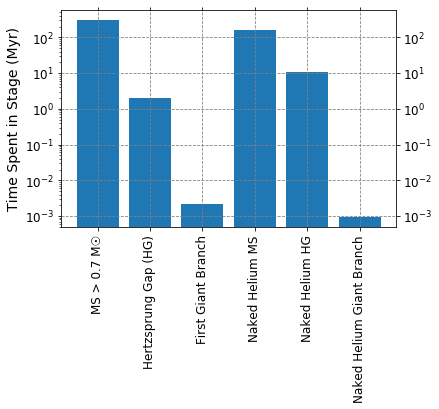} }\\
\multicolumn{2}{c}{(c) Final State: NS+NS}
\end{tabularx}
\caption{Cumulative time spent in a given, populated (e.g. no BH had a MS star < 0.7 $M_{\odot}$ or Red Giant companion) evolutionary state by the secondaries when the primary is a black hole or neutron star. For example, across all 481 BH+BH merging systems, the total time the secondary was a MS star > 0.7 $M_\odot$ was about 500Myr. The typical timescales of merging events (from binary birth to merging) are $\sim 6 Myr$, $\sim 7 Myr$, and $\sim 35 Myr$ for the BH+BH, BH+NS, NS+NS mergers, respectively.}
\label{fig:evolution}
\end{figure*}

\begin{figure}
	\includegraphics[width=\columnwidth]{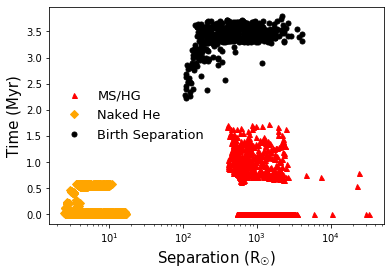}
    \caption{The orbital separations and lifetimes in each evolutionary stage in MYrs for binaries producing BH+BH mergers, labeled by their evolutionary phases: Group of Naked Helium stars (i.e. MS + Hertzsprung Gap) in orange, regular MS + Hertzsprung Gap stars in red, and the birth separation for all 481 binary systems in black. This allows easy visualization for observing how the separation changes with evolutionary time. Note that the other groups can include data for multiple timesteps while the birth separation strictly looks at one specific timestep (the first) of the binary system that will merge as BH+BH.}
    \label{fig:sepbh}
\end{figure}

To represent what type of massive stars have a compact remnant orbiting them, we display a bar graph in Figure \ref{fig:evolution}(a) showing the cumulative time that the secondaries in the BH+BH population spend in each evolutionary state once the primary is a black hole and before the secondary dies. We find that the majority of the time, these black holes orbit around Hydrogen Main-Sequence (MS) or Naked Helium (He) MS stars with considerably less time spent around Hertzsprung Gap (HG) stars. This makes physical sense as the core-burning phases are the longest-lived stages in any star's life cycle. Figure \ref{fig:sepbh} confirms that the orbital separations between the compact remnants and the secondaries are much greater when the secondaries are on the Main-Sequence or Hertzsprung Gap than when they are Naked Helium Stars because most of the mass stripping to produce the He star is done in a CE phase.

For the binaries with MS and Hertzsprung Gap stars, both the separations and the periods, the latter affected by mass loss from the system, are large. Therefore the radial velocities for the secondaries (the only observable stars) are extremely small (see equation \ref{eq:K2}). Figure \ref{fig:bhbhpops}(a) shows the distribution of $\Delta RV_{max}$ for different evolutionary states. In this plot, we find that when the primary is a black hole and the secondary is a living star post-common-envelope (Naked Helium MS + Naked Helium Hertzsprung Gap star), there is a shift in the radial velocity towards much higher values. Although the latter phases are short-lived, they have high RV amplitudes, which will aid in their detection and provide a possible way of observing these non-accreting systems. Not all systems have high $\Delta RV_{max}$ though, because sometimes the limited number of RV measurements were poorly phased or the inclination was high. There is also a gap in the $\Delta RV_{max}$ among the binaries with Naked He companions. This is a result of the evolutionary lifetime of different stages in the Naked He star sequence, and the bottom sequence is mostly composed of short-lived Naked He Hertzsprung Gap stars.

\comment{
\begin{figure}
	\includegraphics[width=\columnwidth]{Evolutionary Time.png}
    \caption{Evolutionary state of the secondary when the primary is a black hole based on the amount of time spent in that specific state. The final end state of these binary systems \textit{will be} a BH+BH merger. The same binary system may be represented more than once in this figure as most systems went from MS/Hertzsprung Gap to Naked Helium Stars.}
    \label{fig:evolutionBHBH}
\end{figure}
}

%\comment{
%\begin{figure}
%	\includegraphics[width=\columnwidth]{BH+BH Scatter.png}
%    \caption{$Light Blue$ represents a binary system in which both stars are living and have large separations, while $orange$ and $red$ represents a system where the primary is a black hole. $Red$, however, is the MS and Hertzsprung Gap stars (see Figure \ref{fig:evolutionBHBH}) while $orange$ is the group of Naked Helium stars.}
%    \label{fig:BHBH}
%\end{figure}
%}

\subsection{BH+NS Mergers}

The progenitors to BH+NS mergers look quite similar to those of the BH+BH  mergers, albeit smaller in number. In particular, we see the same evolutionary states of the secondary when the primary is a black hole (see Figure \ref{fig:evolution}(b)) as well as a similar radial velocity shift (Figure \ref{fig:bhnspops}(a)). The average time before the secondary undergoes core collapse increases from 5.6Myr (BH+BH) to 6.7Myr. Overall, the formation and evolution of these mergers are similar to that of BH+BH mergers. 

%\comment{
%\begin{figure}
%	\includegraphics[width=\columnwidth]{Evolutionary Time (BH+NS).png}
%    \caption{This y-axis log plot shows the evolutionary state of the secondary when the primary is a black hole based on the amount of time spent in that specific state divided by the amount of time the binary systems are dead. In this case, the majority of systems die around 6Myr, so we "kill" all binaries around 10MYr. The final end state of these binary systems \textit{will be} a BH+NS merger. The same binary system may or may not be represented more than once in this figure as most systems went from MS/Hertzsprung Gap to Naked Helium Stars.}
%    \label{fig:evolutionBHNS}
%\end{figure}
%}

\subsection{NS+NS Mergers}

We do find slightly different results when it comes to NS+NS mergers, because these systems have lower masses and longer lifetimes ($\sim35Myr$) before the second star collapses than those systems that contain a black hole. As a result, we find a wider variety of evolutionary stages for the secondaries, including giant branch stars (i.e. First Giant Branch and Naked Helium Star Giant Branch (Figure \ref{fig:evolution}(c)). As with the previous two types of binary systems, these compact object binaries, and eventual mergers, experience a RLOF/CE phase bringing the orbital separation to lower values. Figure \ref{fig:nsnspops}(a) summarizes the distribution of $\Delta RV_{max}$ for these systems. We note that while there is still an upward shift in the radial velocity seen during the Naked Helium sequence, it is significantly lower in amplitude than the previous BH+NS and BH+BH systems, as expected due to the lower masses (see equation \ref{eq:K2}).

\subsection{Non-Mergers}

\begin{figure}
	\includegraphics[width=\columnwidth]{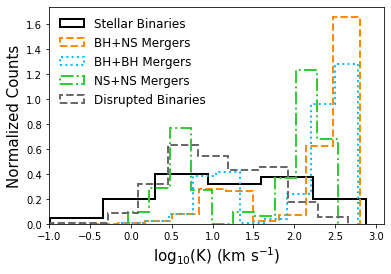}
    \caption{Histograms of the semi-amplitude (K) for all 100,000 binary systems in our default population for each timestep. For the 'Disrupted Binaries,' we stop calculating the RV of the system after disruption, so these RV values correspond to the still-intact binary system. Note that we cut off the x-axis at -1.0, but there were still some 'Stellar Binaries' below this value, albeit very few. Reference Table \ref{tab:kicksims} for the proper definition of 'Stellar Binaries.'}
    \label{fig:semi}
\end{figure}

It is important to know how the systems that do not merge differ from the merging systems in the previous section to compare the detectability of these sources. Figure \ref{fig:semi} illustrates that the general population (i.e. those binaries that are disrupted or do not merge) tend to have much smaller semi-amplitudes, and thus smaller radial velocities. The systems that end up disrupted have the lowest semi-amplitudes on average, as their more loosely bound systems lead neither to high speeds nor the ability to survive mass loss and supernova kicks.

\begin{figure}
	\includegraphics[width=\columnwidth]{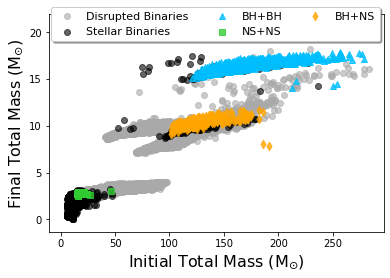}
    \caption{Combined final mass to combined initial mass of each binary system. For those merging systems, this is just the final mass of the remnant. We note that there are many stellar binaries that overlap with the BH+BH merger systems in the top concentration.}
    \label{fig:masses}
\end{figure}

We are also interested whether any system, merging or non-merging, can produce BH masses as high as the extremes detected by LIGO.
Figure \ref{fig:masses} shows what combined initial mass of the system translates to combined final mass of the system. As expected, the higher the initial mass, the higher the final mass, which tends to produce BH+BH mergers. However, it is clear from this figure that these systems do not produce the most massive LIGO/Virgo sources (>30$M_{\odot}$). This could be because the default maximum mass imposed in COSMIC is 150$M_{\odot}$, and in combination with mass loss, these systems are simply not created. As noted by \citet{Rodriguez_2018}, dynamical processes and enhanced interaction rates in clusters may therefore be critical in creating black holes at masses $>$ 50 M$_{\odot}$. 

\section{Testing Parameter Space}
\label{sec:params}

In Section~\ref{sec:sims}, we generated a population with COSMIC using all of the default parameters; the configuration file records all flags, filters, sampling methods, and convergence data.  However, we are also interested in how changes in certain parameters affected the predicted properties of the binary population. As a result, we ran more simulations of the same binary population size with different prescriptions to see the influence on the data. Note that there are a few parameters that cannot be changed that are embedded into COSMIC's multidim algorithm, such as the IMF.

In particular, we explore the results of (1) changing the distribution of natal kicks and (2) changing the metallicity of the binary population. Table \ref{tab:kicksims} shows the final end states for all of the binary systems in the different populations, including the default population for easy comparison. While we set all of the simulations to have 100,000 binary systems, sometimes one fewer binary system was generated even though we use the multidim sampler where the binary fraction is a parameter and is not user-specified. Despite private conversation with the creator of COSMIC, we do not know the cause, and refer the reader to section 2.1.2 of \cite{Breivik2020} as a possible source.

\begin{figure*}
\begin{tabularx}{1\textwidth}{cc}
\centering
  \includegraphics[width=85mm]{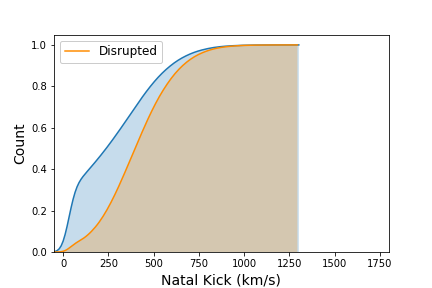} &   \includegraphics[width=85mm]{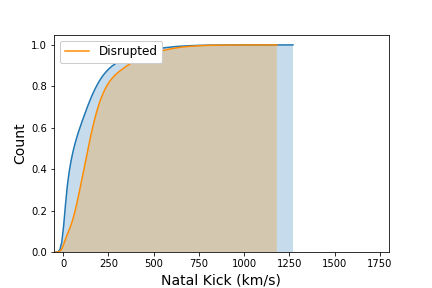} \\
(a) Standard & (b) Eq. 1 of \citeauthor{Giacobbo2020} \\[6pt]
 \includegraphics[width = 85mm]{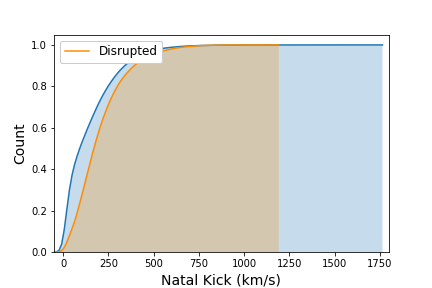} &   \includegraphics[width = 85mm]{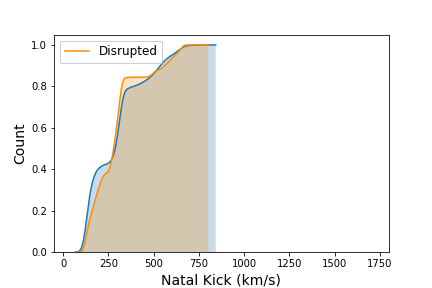} \\
(c) Eq. 2 of \citeauthor{Giacobbo2020} & (d) Eq. 1 of \citeauthor{Bray2016} \\[6pt]
\includegraphics[width = 85mm]{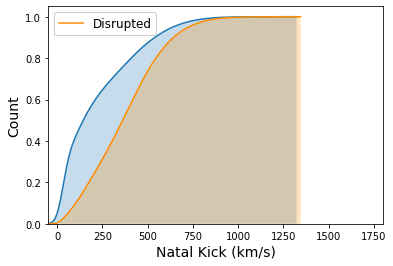} & \includegraphics[width = 85mm]{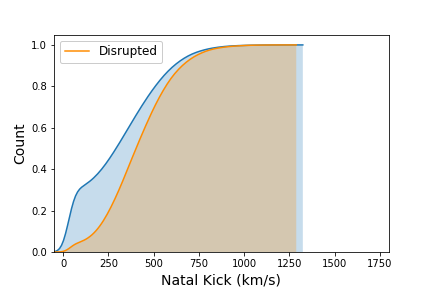}\\
(e) Z = 0.01 & (f) Z = 0.03
\end{tabularx}
\caption{Cumulative, normalized histograms summarizing how the distribution of natal kicks change as certain parameters are modified.  Systems that were disrupted by the kick are shown in orange, while the blue histogram represents those that remain binaries. The most notable change is when equation one of \citeauthor{Bray2016} is used, which is where we also see the highest number of disruptions. However, the other kick prescriptions see a decrease in the total number of disruptions compared to our default population.}
\label{fig:kickpops}
\end{figure*}

The kick prescription affects how many binaries are disrupted at the first or second supernova, which in turn affects the number of gravitational wave sources we can expect. We consider a total of four simulations. The first used the default kickflag given by COSMIC, in which kicks are drawn from a bimodal distribution. Standard core-collapse supernovae (one mode) draw their kicks from a Maxwellian with a sigma value of 265 km s$^{-1}$, given by \citet{Hobbs2005}. Two other simulations we ran include the revised natal kick distributions given by Equation 1 and Equation 2 in \cite{Giacobbo2020}. These natal kicks are still drawn from a sigma value of $265 {\rm km s}^{-1}$, but are then scaled by the ejecta mass and remnant mass or just the ejecta mass, respectively. Finally, we ran a simulation utilizing Equation 1 of \citet{Bray2016}. Figure \ref{fig:kickpops}(a-d) compares the natal kick distributions given by these four prescriptions for both the systems that remain gravitationally bound and those that are disrupted. Unsurprisingly, we see that the non-disruptive kicks are generally lower in magnitude compared to those that unbind the system, but some systems can withstand high kick velocities.  

%In addition to running the same analysis in Section \ref{sec:respop}, we also plot the magnitude of the natal kicks given the different prescriptions used for a given population.

Metallicity is well known to affect a star's luminosity, effective temperature/colour, and mass-loss rate via stellar winds \citep[e.g.,][]{Mapelli2013}. All of these parameters contribute greatly to the observed properties of the star and to its evolution. For example, the size of its carbon-oxygen core, that will in turn affect the final remnant of a star \citep[e.g.,][]{Spera_2015}. Because of this, it is very important to consider the metallicity when studying the formation of compact binaries. In addition to the solar metallicity simulation ran in Section~\ref{sec:sims}, in which had the metallicity of the binary population set to the solar value, $Z = 0.02$, we also ran simulations with the metallicity set to $Z = 0.03$ (more metal-rich) and $Z = 0.01$ (more metal-poor). 

\begin{figure*}
\begin{tabularx}{1\textwidth}{cc}
\centering
  \includegraphics[width=85mm]{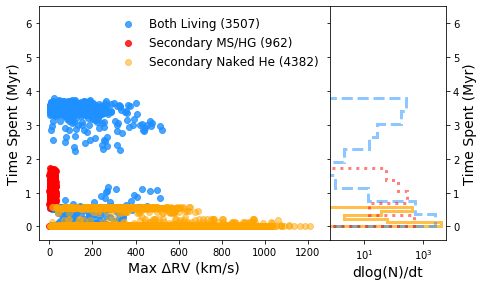} &   \includegraphics[width=85mm]{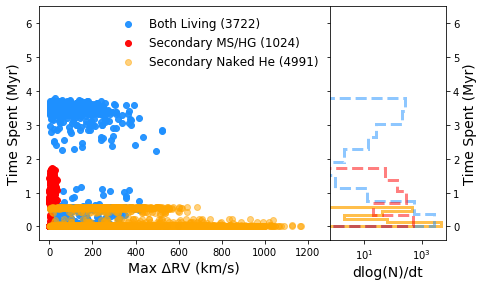} \\
(a) Standard: Maxwellian with $\sigma = 256$km s$^{-1}$  & (b) (Kick = $-1$) Eq. 1 of \citeauthor{Giacobbo2020}: v$_{\rm kick}$ $= f_{\rm H05} \frac{m_{\rm ej} \langle m_{\rm NS} \rangle}{\langle m_{\rm ej} \rangle m_{\rm rem}}$ \\[6pt]
 \includegraphics[width = 85mm]{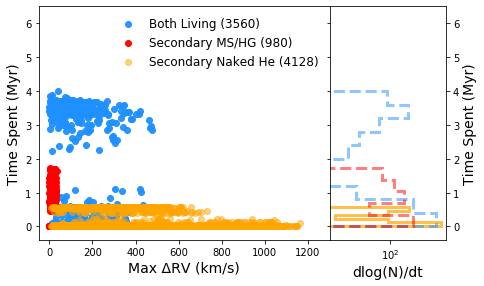} &   \includegraphics[width = 85mm]{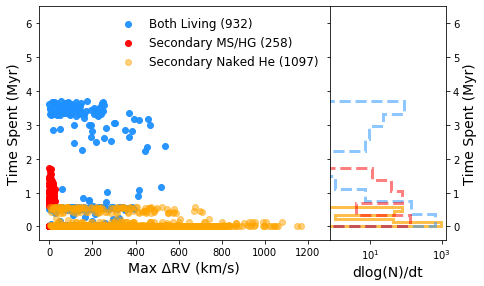} \\
(c) (Kick = $-2$) Eq. 2 of \citeauthor{Giacobbo2020}: v$_{\rm kick}$ $= f_{\rm H05} \frac{m_{\rm ej}}{\langle m_{\rm ej} \rangle}$ & (d) (Kick = $-3$) Eq. 1 of \citeauthor{Bray2016}: v$_{\rm kick}$ $= \alpha \frac{M_{\rm ejecta}}{M_{\rm remnant}}+\beta$ \\[6pt]
\multicolumn{2}{c}{\includegraphics[width=85mm]{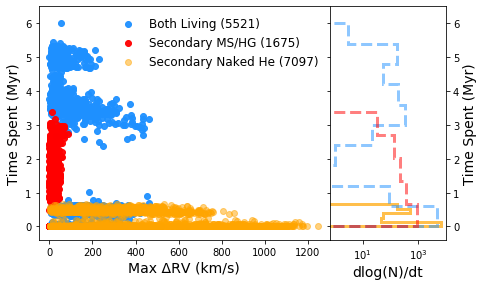} }\\
\multicolumn{2}{c}{(e) Z = 0.01}
\end{tabularx}
\caption{The distribution of Max $\Delta$RV for systems with at least one living star before they end as BH+BH mergers for populations with different kick parameters and metallicities. Blue represents a binary system in which both stars are living, while orange and red represents a system where the primary is a black hole. Red points represent MS and Hertzsprung Gap stars (see Figure \ref{fig:evolution}(a)) while orange is the group of Naked Helium stars. The most obvious effect of changing the metallicity is the predicted absence of BH+BH mergers for the metal-rich population (see Table \ref{tab:kicksims}), which is why there is no figure displayed for Z = 0.03. Note that those points close to 0.0 are not all exactly equal to zero. These points are a result of Naked He Hertzsprung Gap star and Naked He MS star with rapid timesteps generated by COSMIC, indicating the start and end of RLOF/CE.}
\label{fig:bhbhpops}
\end{figure*}

Figure~\ref{fig:bhbhpops} shows the Max $\Delta$RV distribution for the systems that merged as BH+BH binaries for the default parameter run and for the changed kick and metallicity runs. The duration of the phase where a Max $\Delta$RV for a system was derived is shown on the y-axis. Appendix \ref{appendix:graphs} shows the analogous plots for the BH+NS and NS+NS mergers (\ref{fig:bhnspops}, \ref{fig:nsnspops}). As expected, regardless of choice of metallicity or kick parameters, the systems with a Naked HE star + BH show the largest Max $\Delta$RV. When the systems have two living stars, they can also have large Max $\Delta$RV, but before the envelope of the second star is stripped, BHs orbiting MS stars are the most difficult to detect with RVs.

However, there are large changes in the number and lifetimes of systems in different phases. The most dramatic example is the complete absence of BH+BH and BH+NS mergers, and only a few NS+NS mergers at Z=0.03. These stars lose a lot of mass throughout their lifetimes because of the prescription for the strength of stellar winds as a function of metallicity; therefore, they are no longer massive enough to form BHs at the end of their lives. \cite{Heger_2003}. In contrast, our more metal-poor population saw the highest number of BH+BH mergers, but one of the lowest numbers of BH+NS mergers. This population also had less massive binaries (with $\sim$ 40 M$_{\odot}$ stars) end up as BH+BH mergers, which led to a longer amount of time spent in the MS stage. This is seen in Figure \ref{fig:bhbhpops}(e) where we have two distinct concentrations when both stars are living. The changes in the number of sources for our kick prescriptions match nicely with Figure \ref{fig:kickpops}. Figures (b) and (c) see the highest number of all mergers, due to the inherently low velocity kicks for non-disruptions. Figure (d) see some of the lowest numbers, which is a result of the very high natal kicks (at least 100 ${\rm kms}^{-1}$) present in the population. Besides changes seen with the number of mergers, the formation mechanisms and radial velocities of these systems are similar to our initial population.

\begin{table*}{Comparison of Final End States for Populations with Different Parameters}
    \centering
    \caption{A summary table of the number of sources in our different populations. 'Stellar Binaries' are simply those systems that remain gravitationally bound, but do not merge in the amount of time that the user has specified for evolution to occur. While we use "stellar" to describe such systems, the final remnants of the stars may be a compact object depending on the initial mass (i.e. an initial 15 $M_{\odot}$ will not be a star in 13.7 GYr). In our example, this maximum evolutionary time is the lifetime of the Universe. 'Disrupted Binaries' (a subset of the 'Stellar Binaries') include those binary systems that were disrupted only due to a supernova, and thus do not merge. 'Stellar Mergers' refers to systems that merged in various stages of evolution that were not black holes or neutron stars.}
	\label{tab:kicksims}
    \begin{tabularx}{0.9\textwidth}{c r r r r c c c}
    \toprule
Population & Total \# of Systems & \# BH+BH & \# NS+BH & \# NS+NS & \# Stellar Binaries & \# Disrupted Binaries & \# Stellar Mergers \\ \midrule
Default & 99999 & 481 & 198 & 87 & 49200 & 28943 & 50033 \\
Kick = $-1$ & 100000 & 512 & 214 & 147 & 42047 & 18272 & 57080 \\
Kick = $-2$ & 100000 & 490 & 267 & 141 & 43329 & 20312 & 55773 \\
Kick = $-3$ & 100000 & 129 & 68 & 17 & 50589 & 33266 & 49197 \\ 
Z = 0.01 & 100000 & 771 & 80 & 118 & 49753 & 28735 & 49278 \\
Z = 0.03 & 99992 & 0 & 0 & 68 & 48192 & 27556 & 51732 \\
\bottomrule
    \end{tabularx}
\end{table*}

\comment{\multicolumn{2}{c}{\includegraphics[width=65mm]{it} }\\
\multicolumn{2}{c}{(e) fifth}}

\section{Discussion}

We have shown in Section~\ref{sec:respop} the Max $\Delta$RV expected for systems with two massive stars as they evolve past their main-sequence lifetimes, through the supernova stage, and possibly end up as mergers. These changes in RV, especially for the systems with Naked He stars, are large enough to be easily detected in few-epoch surveys with $\sim$ 10 km s$^{-1}$ velocity precision, such as LAMOST and MWM, provided that the stars are targeted for observations. 

However, the exact number and configuration of the systems depends on the adopted parameters and on the underlying assumptions of the evolution of stars in COSMIC. We highlight two concerns here related to the use of the \cite{Fryer2012} prescription for predicting the remnant masses produced by core-collapse SNe. First, fallback that can occur onto the forming remnant fills in the mass gap between BHs and NSs, which is in contradiction to the current observational evidence. The most recent gravitational wave catalogue from the LIGO/Virgo Collaboration \citep[e.g.,][]{AbbottGWTC3} still see no remnants in this gap, with the highest probability of a candidate in the gap being only 13$\%$. Further, no modern theoretical models predict significant fallback; stars really either explode, or they do not. Second, \citet{Patton_2020} point out that the relationship between remnant mass and main-sequence mass/carbon-oxygen core mass is not monotonic. They evolve a suite of CO cores with a range of composition and mass to explosion and find that only some cores exploded as SNe. The rest imploded directly to black holes, creating more massive remnants than previously expected. While these concerns will affect the details of the predictions, the main conclusions about the expected RV range and evolutionary state of companions remain. Indeed, we can hope that extensive RV observations and comparison with different remnant mass prescriptions can illuminate the explodability of massive stars.

Further, we have focused here on radial velocity detection, which has the advantage of being able to probe distant systems, including massive stars in lower-metallicity environments such as the Magellanic Clouds. However, only a limited number of stars will be observed over at least a few epochs in current and planned surveys, while other methods such as astrometry also provides a way to detect Galactic black holes \citep[e.g.,][]{Lu2016, Breivik_2017}. 

Finally, \cite{Wik2020} estimated how many non-interacting binary black holes are within the Milky Way, particularly those that are observable astrometrically by {\it Gaia} and spectroscopically by LAMOST. They find very similar results to ours, despite using different BPS models. Our evolutionary states tend to match their {\it Gaia} sample, where the living stars are luminous and massive, with black holes having an enhanced probability of being found around helium stars. They also find that systems with small separations contain a helium star and systems with large separations contain a main-sequence star that has not yet interacted with its companion. These results agree with our analysis as well. They predict a large fraction of LAMOST stars with a detectable non-interacting BH companion will be main-sequence stars, because they do not confine their analysis to systems where both stars are high mass.

\section{Conclusions}

%The last numbered section should briefly summarise what has been done, and describe
%the final conclusions which the authors draw from their work.

We investigate the observability of progenitor systems to BH+BH, BH+NS,
and NS+NS mergers in few-epoch RV data when at least one object is still a star and how the kick distribution and the metallicity affect the distribution of system parameters. We focus on the observing strategies used by recent and current surveys such as APOGEE, MWM, and
LAMOST. We find that the highest velocity dispersion comes from systems with an observable Wolf-Rayet star or hot subdwarf (Naked Helium type star) and a BH or NS due to in-spiral (like the detected sources from LIGO/Virgo) likely resulting from CE and RLOF, bringing the orbital separation lower and lower. On the other hand, there are rather low values for $\Delta RV_{max}$ in systems that contain a MS, HG, or Giant Branch (GB) (in the case of NS+NS mergers) star in comparison to systems with Naked Helium type stars. Furthermore, when we change certain parameters known to affect binary star evolution, we still observe a shift in these respective categories, but obtain a different numbers of sources. Specifically, changes in the kick models have modest effects as does reducing the metallicity to half solar. The only really dramatic change comes from models with 50\% higher metallicity where black holes essentially no longer form. All in all, surveys such as Milky Way Mapper will observe \textit{N} number of systems, and in combining the results of all the previous plots, we hope to refine how many stars we can expect of each type (i.e. MS, Hertzsprung Gap, Naked Helium) and what $\Delta RV_{max}$ they should have. From this data, we can learn what RV sensitivity is required for future surveys to achieve to be able to acquire such observations.

\section*{Acknowledgements}

%The Acknowledgements section is not numbered. Here you can thank %helpful
%colleagues, acknowledge funding agencies, telescopes and facilities used etc.

This paper is dedicated to the loving memory of Jean A. Banner-Carroll.

We thank Katie Breivik for invaluable technical assistance and advice. We also thank Rachel Patton for useful discussion and Chris Kochanek as well as Tharindu Jayasinghe for critical comments on early drafts of this paper.

We would like to acknowledge the land that The Ohio State University occupies is the ancestral and contemporary territory of the Shawnee, Potawatomi, Delaware, Miami, Peoria, Seneca, Wyandotte, Ojibwe and Cherokee peoples. Specifically, the university resides on land ceded in the 1795 Treaty of Greeneville and the forced removal of tribes through the Indian Removal Act of 1830. As members of a land grant institution, we want to honor the resiliency of these tribal nations and recognize the historical contexts that has and continues to affect the Indigenous peoples of this land.

%%%%%%%%%%%%%%%%%%%%%%%%%%%%%%%%%%%%%%%%%%%%%%%%%%
\section*{Data Availability}

COSMIC is a publicly available code and the simulations can be generated using the described parameters. If the complete data used to create the figures are needed, the data will be supplied upon reasonable request.



\bibliographystyle{mnras}
\bibliography{main} % if your bibtex file is called example.bib

%\begin{thebibliography}{99}
%\end{thebibliography}

%%%%%%%%%%%%%%%%%%%%%%%%%%%%%%%%%%%%%%%%%%%%%%%%%%

%%%%%%%%%%%%%%%%% APPENDICES %%%%%%%%%%%%%%%%%%%%%

\begin{appendices}

\section{Graph Appendix}
\label{appendix:graphs}

\begin{figure*}
\begin{tabularx}{1\textwidth}{cc}
\centering
  \includegraphics[width=85mm]{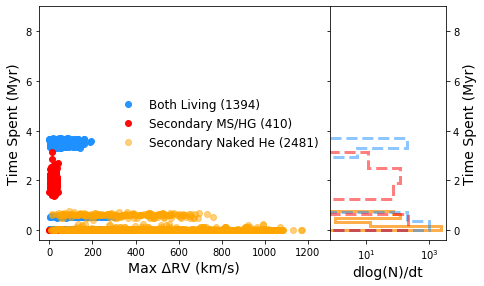} &   \includegraphics[width=85mm]{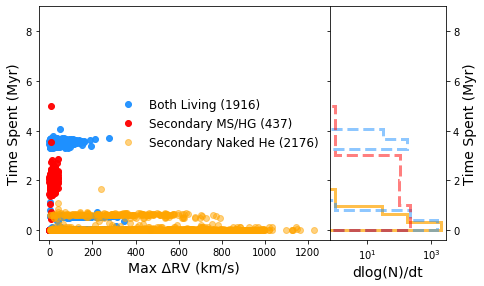} \\
(a) Standard: Maxwellian with $\sigma = 256$km s$^{-1}$ & (b) (Kick = $-1$) Eq. 1 \citep{Giacobbo2020}: v\_{kick} $= f_{H05} \frac{m_{ej} \langle m_{NS} \rangle}{\langle m_{ej} \rangle m_{rem}}$ \\[6pt]
 \includegraphics[width = 85mm]{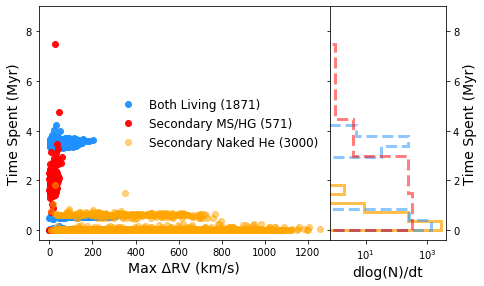} &   \includegraphics[width = 85mm]{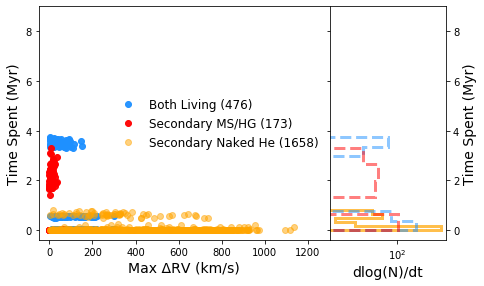} \\
(c) (Kick = $-2$) Eq. 2 \citep{Giacobbo2020}: v\_{kick} $= f_{H05} \frac{m_{ej}}{\langle m_{ej} \rangle}$ & (d) (Kick = $-3$) Eq. 1 \citep{Bray2016}: v\_{kick} $= \alpha \frac{M_{ejecta}}{M_{remnant}}+\beta$ \\[6pt]
\multicolumn{2}{c}{\includegraphics[width=85mm]{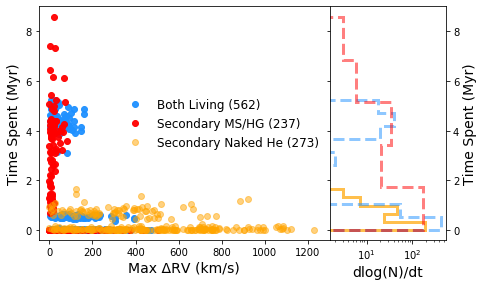} }\\
\multicolumn{2}{c}{(e) Z = 0.01}
\end{tabularx}
\caption{As in Figure \ref{fig:bhbhpops}, but for BH+NS.}
\label{fig:bhnspops}
\end{figure*}

\begin{figure*}
\begin{tabularx}{1\textwidth}{cc}
\centering
  \includegraphics[width=85mm]{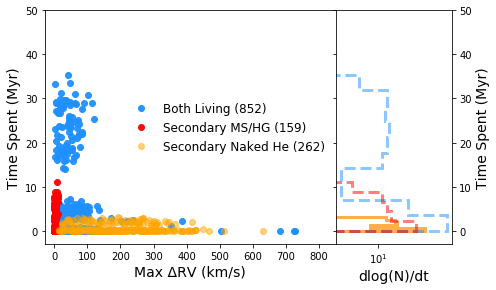} &   \includegraphics[width=85mm]{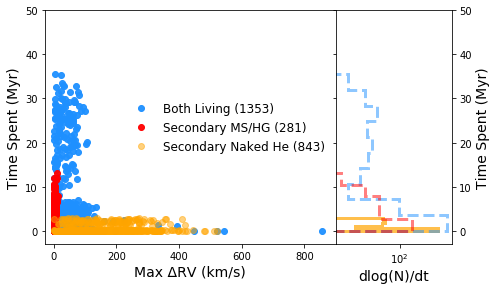} \\
(a) Standard: Maxwellian with $\sigma = 256$km s$^{-1}$ & (b) (Kick = $-1$) Eq. 1 \citep{Giacobbo2020}: v\_{kick} $= f_{H05} \frac{m_{ej} \langle m_{NS} \rangle}{\langle m_{ej} \rangle m_{rem}}$ \\[6pt]
 \includegraphics[width = 85mm]{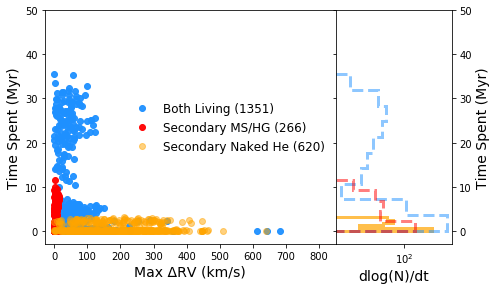} &   \includegraphics[width = 85mm]{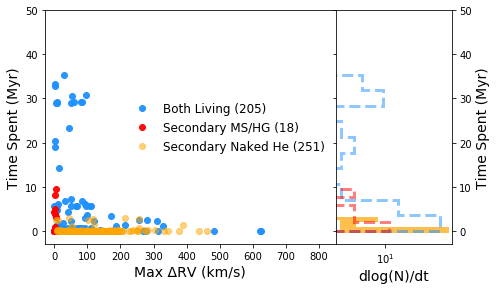} \\
(c) (Kick = $-2$) Eq. 2 \citep{Giacobbo2020}: v\_{kick} $= f_{H05} \frac{m_{ej}}{\langle m_{ej} \rangle}$ & (d) (Kick = $-3$) Eq. 1 \citep{Bray2016}: v\_{kick} $= \alpha \frac{M_{ejecta}}{M_{remnant}}+\beta$ \\[6pt]
\includegraphics[width = 85mm]{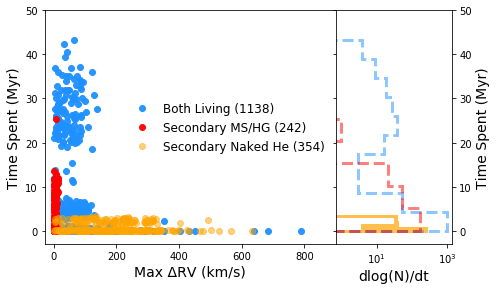} & \includegraphics[width = 85mm]{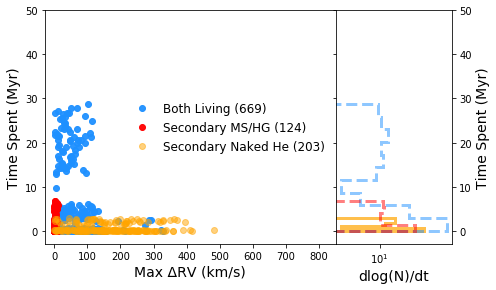}\\
(e) Z = 0.01 & (f) Z = 0.03 
\end{tabularx}
\caption{As in Figure \ref{fig:bhbhpops}, but for NS+NS.}
\label{fig:nsnspops}
\end{figure*}

Figure~\ref{fig:bhnspops} and Figure~\ref{fig:nsnspops} show the RV distributions for BH+NS and NS+NS mergers.

\end{appendices}

%%%%%%%%%%%%%%%%%%%%%%%%%%%%%%%%%%%%%%%%%%%%%%%%%%

% Don't change these lines
\bsp	% typesetting comment
\label{lastpage}
\end{document}